 \documentclass[pmlr,twocolumn,10pt]{jmlr} 

\usepackage{booktabs}
\usepackage[load-configurations=version-1]{siunitx} 
\usepackage{bbm}
\usepackage{bm}
\usepackage{mathtools}
\usepackage{float}
\usepackage{wrapfig}

\newcommand{\equal}[1]{{\hypersetup{linkcolor=black}\thanks{#1}}}

\theorembodyfont{\upshape}
\theoremheaderfont{\scshape}
\theorempostheader{:}
\theoremsep{\newline}

\title[Efficient self supervision]{Resource and data efficient self supervised learning}

\author{%
\Name{Ozan Ciga}\equal{These authors contributed equally} \Email{ozan.ciga@mail.utoronto.ca}\\
\addr Department of Medical Biophysics, University of Toronto, Canada
\AND
\Name{Tony Xu}\footnotemark[1] \Email{tony.xu@alumni.ubc.ca}\\
\addr Department of Electrical and Computer Engineering, University of British Columbia, Canada
\AND
\Name{Anne L. Martel} \Email{a.martel@utoronto.ca}\\
\addr Physical Sciences, Sunnybrook Research Institute, Toronto, Canada
}

\begin{document}

\maketitle

\begin{abstract}
  We investigate the utility of pretraining by contrastive self supervised learning on both natural-scene and medical imaging datasets when the unlabeled dataset size is small, or when the diversity within the unlabeled set does not lead to better representations. We use a two step approach which is analogous to supervised training with ImageNet initialization, where we pretrain networks that are already pretrained on ImageNet dataset to improve downstream task performance on the domain of interest. To improve the speed of convergence and the overall performance, we propose weight scaling and filter selection methods prior to second step of pretraining. We demonstrate the utility of this approach on three popular contrastive techniques, namely SimCLR, SWaV and BYOL. Benefits of double pretraining include better performance, faster convergence, ability to train with smaller batch sizes and smaller image dimensions with negligible differences in performance. We hope our work helps democratize self-supervision by enabling researchers to fine-tune models without requiring large clusters or long training times. 
\end{abstract}

\section{Introduction}
The ability to infer meaning and identify patterns from unstructured data, or unsupervised learning, has been a goal of machine learning researchers which predates the advancements in deep learning \citep{xu2005survey}. A generalized data-independent framework for learning features and patterns from visual input became possible with the advent of self-supervised techniques. The early examples of these techniques commonly employ the supervised learning objective and generate the supervision signal from the raw input. More recently, self-supervised methods based on contrastive learning have consistently outperformed their predecessors in various vision tasks and achieved state-of-the-art results on the popular ImageNet classification benchmark \citep{tian2019contrastive, he2020momentum, chen2020simple}. These methods are exhaustively pretrained on large amounts of data and are deployed for further use in downstream tasks similar to transfer learning based on supervised training (e.g., ImageNet pretraining).

While current techniques are not overly complex in nature, they require access to hardware such as multiple GPUs or TPUs for large batch training for a large number of epochs that is not readily available to most researchers and practitioners. Furthermore, in most domains, acquiring large amounts of unlabeled data is not straightforward and may be subject to regulation (e.g., medical images) \citep{srinidhi2020deep}. Finally, the quality of learned representations is dependent on identifying visual differences between contrasted samples. This may be challenging in complex tasks where most images do not exhibit high diversity, such as cancer detection from large resolution biopsy images where a single cell tumor may change the decision outcome. In such cases, learning directly from raw data generally encodes noisy features which are not superior to a randomly initialized network in downstream tasks \citep{ciga2020self}.

This paper proposes a simple transfer learning approach based on pretraining an already pretrained network on a new domain to improve performance on downstream tasks. We demonstrate the utility of this approach on three contrastive self-supervised methods. Using each method, we first pretrain a Resnet50 using natural-scene images from ImageNet ILSVRC-2012 dataset \citep{ILSVRC15}. Learned representations are then fine-tuned by a second stage of pretraining on data from a second domain. In this work, we select the second stage data according to the downstream task. For example, if the task is to identify different brands of cars from images, the second stage dataset is a set of unlabeled car images. We verify the efficacy of this procedure compared to both self-supervised and ImageNet initializations in multiple domains, regardless of the similarity between first and second stage datasets (e.g., natural-scene images to medical images). This approach improves downstream task performance even when there is limited data for pretraining, which allows for shorter training times, has better downstream task performance when trained for the same amount as pretraining from scratch, and is able to work with smaller batch sizes and can be pretrained using smaller images, mitigating the hardware requirements.

\section{Related work}

Self-supervised learning for images has become more popular in recent years due to its promise of alleviating the requirements for labeled data. Its aim is to learn latent-space representations through a learning objective without human annotations which can be used in downstream supervised tasks such as classification. Early context-based methods exploit the spatial regularity in images by altering an image in total or in part to train a network to recover proper arrangement, alignment, or orientation from the modified image \citep{doersch2015unsupervised, noroozi2016unsupervised, gidaris2018unsupervised}. While these techniques are superior to their contemporary unsupervised counterparts, they are based on handcrafted pretext tasks, which can bias and limit the learning. 

Contrastive self-supervised methods replace such heuristically determined pretext tasks by comparing multiple images to each other and assigning each pair into so-called positive and negative classes. These methods assume that two augmented versions of the same image are positive, whereas each pair with distinct samples is considered negative. The learning objective is then to bring latent-space representations of positive pairs closer in some metric space and to push the negative pairs' representations apart. Negative instances can be stored in a dynamic memory bank per training instance to avoid recomputing feature vectors \citep{wu2018unsupervised, bachman2019learning, he2020momentum, chen2020big}. Augmentations such as scaling, affine transformations, or adjusting color properties of images are widely used in contrastive learning to avoid trivial solutions and to improve robustness of learned features \citep{tian2019contrastive, henaff2020data, misra2020self, chen2020simple}. A recent class of methods rely on large minibatch training where each sample is a negative for the other samples in the batch except for the sample's augmented view. These methods apply different architectural and design choices such as temperature based loss functions, exponential moving average of the model weights, and comparing prototype vectors instead of raw latent representations to improve the performance on downstream tasks \citep{chen2020simple, grill2020bootstrap, caron2020unsupervised}.

\begin{figure*}
     \centering
     \includegraphics[width=1\textwidth]{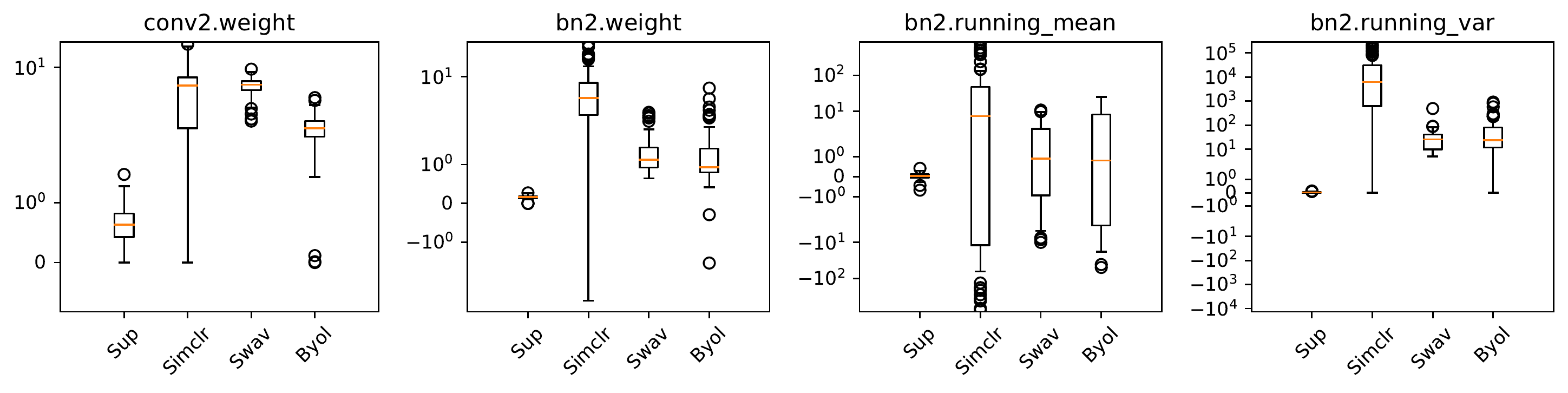}
        \caption{Comparison of filter weight distributions for different methods of pretraining. ``conv" weights are summarized by taking the Frobenius norm for each layer, whereas other quantities represent raw numbers. Resnet50 trained with supervision from ImageNet dataset (Sup) has only 0.5\% of its filters with Frobenius norm above 1, whereas it is 96\% for both SimCLR and SwaV, and 68\% for BYOL.}\label{fig:compare_weights}
\end{figure*}

Transfer learning, the ability to utilize information encoded in a network to improve performance on another task \citep{pan2009survey, weiss2016survey}, is widely used in variety of machine learning tasks, including computer vision \citep{He2016}, natural language processing \citep{radford2018improving, devlin2018bert}, and speech recognition \citep{kunze2017transfer}. In a broad sense, domain adaptation approaches, where transferring the learning of a source with the same label space as the target, can be considered a form of transfer learning \citep{ganin2015unsupervised, shu2018dirt}. In addition to the architectural, conceptual, and methodological modifications \citep{shimodaira2000improving, ganin2015unsupervised, maicas2018training, saito2017asymmetric, shu2018dirt}, weight regularization has been also employed to achieve transfer learning. Approaches such as hypothesis transfer learning \citep{kuzborskij2013stability, kuzborskij2017fast} and multi-model knowledge transfer \citep{tommasi2013learning}, aim to transfer knowledge in a fast and a stable manner by limiting the weight drift with a penalty term on the difference between the target and source domain weights, or by introducing a penalty term which resembles the moving average of weights (e.g., $(1-\alpha)\left\lVert\ \beta \right\rVert_2 ^2 + \alpha \left\lVert\ \beta - \hat{\beta} \right\rVert_2 ^2$, where $\beta$ is the target weights, $\hat{\beta}$ is the source weights, and $\alpha$ is a nonzero momentum parameter). Recently, \cite{takada2020transfer} showed that $\ell_1$ regularization can be used for transfer learning, which simultaneously enforces sparsity while dampening down all the weights. 

In the context of current machine learning techniques, a network can encode useful information by pretraining, which can be either supervised or unsupervised. The resulting pretrained network is then used as initialization for use in downstream tasks for further training. Recently, \cite{dontstoppretraining2020} proposed a two stage pretraining approach for natural language processing tasks. In the first stage, a general language model is trained which is fine tuned in a subsequent stage for domain-specific tasks. The authors showed that this two tiered approach can be used for transfer learning in downstream tasks where the domain of interest is similar or same as the data used for second stage of training.

\section{The method}\label{sec:method}

\paragraph{Summary} Our method is analogous to supervised training with ImageNet initialization, where we use the Resnet50 pretrained on images from the ILSVRC-2012 as initialization. We use the same pretraining method used to obtain the initialization weights and pretrain a second time with images from the domain of interest. We found this straightforward approach sometimes requires a longer second stage of pretraining and propose an adjustment to the originally pretrained weights, as described below. The modifications described below do not lead to any performance gains for pretraining from scratch, as most networks using Xavier \citep{glorot2010understanding} or He normalization \citep{he2015delving} by default eschew problems such as dead filters or exploding weights. 


\paragraph{Exploding weights} We compare the convolutional filter weights and batch normalization parameters and statistics of supervised and self-supervised networks that are trained on ImageNet. We find that the magnitudes of parameters for the supervised network are significantly smaller (Fig. \ref{fig:compare_weights}), likely due to the small weight decay employed in contrastive techniques compared to supervised training \citep{chen2020simple, caron2020unsupervised}. We argue this may lead to optimization issues for further pretraining tasks, and scale weights and batch normalization parameters by the Frobenius norm of each corresponding layer, if the norm is greater than 1. We illustrate the scaling operation using the batch normalization equation. Assume the input $x$ is convolved using the function $f_c(\cdot)$ at layer $i$:  $\frac{f_c(x)-\mu_{rm}}{\sqrt{\sigma^2_{rv} + \epsilon}} * \gamma + \beta$, where $\mu_{rm}, \sigma^2_{rv}$ are the running mean and variance, and $\gamma, \beta$ are the affine parameters for batch normalization at $i$, and $\epsilon$ is a small number for numerical stability. Prior to second pretraining step, we calculate the Frobenius (matrix) norm $s$ for the layer. We scale each component by a function of $s$, which modifies the previous equation as $\frac{f_c(x)/\sqrt{s}-\mu_{rm}/\sqrt{s}}{\sqrt{\sigma^2_{rv}/s^2 + \epsilon}} * \gamma/\sqrt{s} + \beta$, so that the output of batch normalization is the same, however the numerical values of each parameter are reduced by a factor of $\sqrt{s}$ or $s^2$. We use the square root to be able to scale the convolutional filters, the running mean, and the affine batch normalization parameters simultaneously. We scale the convolutional filter weights instead of x (the input), as the convolution is linear.

Prior to using Frobenius norm scaling, we also experimented with a single universal scaling term that was used to scale all weights by the same amount; however, this did not improve the two-step pretraining. Similarly, using a larger weight decay to dampen weights also led to suboptimal performance. 

\begin{figure*}
     \centering
     \includegraphics[width=1\textwidth]{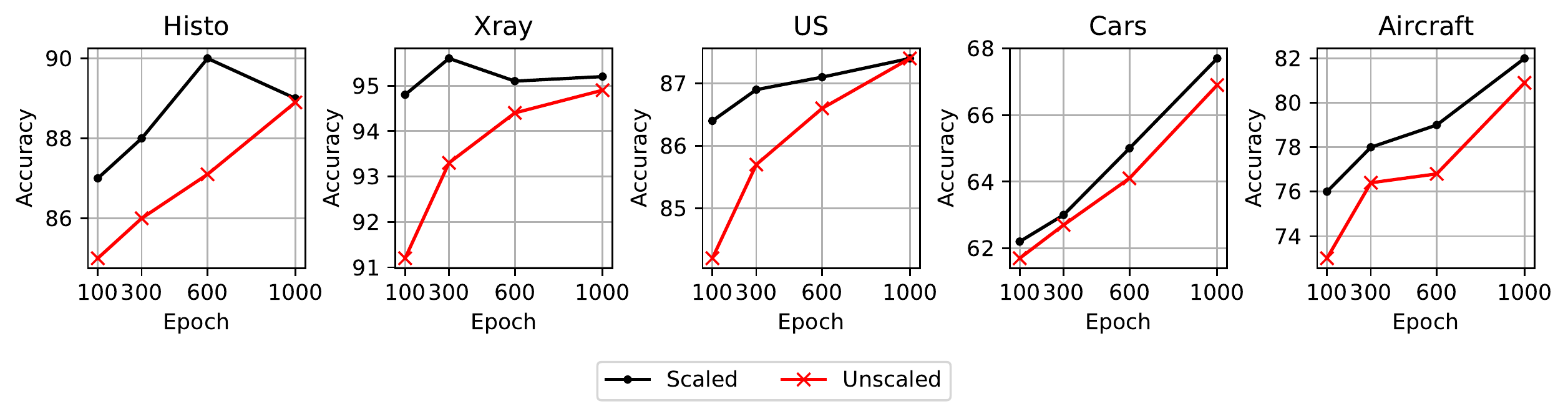}
    \caption{The impact of weight scaling prior to SimCLR pretraining on convergence for the five randomly selected datasets. The values show the accuracy at 100, 300, 600, and 1000$^{th}$ epochs for both unscaled and scaled initializations. The difference between downstream accuracy values diminish as the pretraining progresses.}\label{fig:faster_scaling}
\end{figure*}

We found that scaling led to faster convergence. We obtained [0.7\%, 5.5\%] improvement over straightforward training in validation experiments without weight scaling when we only pretrained for 100 epochs at the second stage. The difference between scaled and non-scaled initializations has diminished as the networks were pretrained for more epochs, reaching virtually the same validation accuracy (see Fig. \ref{fig:faster_scaling}).

\begin{table}
\caption{Comparison of different filter selection methods.}\label{wrap-tab:dead_filters}
\begin{tabular}{cccccc}\\\toprule  
Method & Histo & Xray & US & Cars & Aircraft \\ \midrule 
Baseline & \textbf{89.3} & 88.8 & 95.0 & 64.3 & 81.5 \\
Random & 87.1 & 90.1 & 94.9 & 67.1 & 81.8 \\ 
Copy & \textbf{89.3} & \textbf{90.8} & \textbf{95.5}  & \textbf{67.6}  & \textbf{82.3} \\ \bottomrule
\end{tabular}
\end{table}

\paragraph{Dead filters} We also found some convolutional filters at each layer of the Resnet50 network having Frobenius norms below 0.1 ($\sim$ 1\%, 0.2\% and 1\% for SimCLR, SwaV and BYOL, respectively), which leads to zero response for that filter. Similar to the dying ReLU problem, this may cause gradient updates to be zero for that filter, leading to capacity underuse. Replacing these filters by randomly initialized filters ameliorates the problem; however, we heuristically found randomly copying filters at the same layer that have Frobenius norms above 0.1 and replacing them with these dead filters to work better. The experiments comparing different filter selection schemes for the SimCLR pretraining method validated on five datasets (see Section \ref{sec:experiments}) are shown in Table \ref{wrap-tab:dead_filters}.


\subsection{Contrastive techniques}\label{sec:method_simclr}

In this section, we give a brief overview of the three contrastive self-supervised methods that are used for the two-step pretraining. As each method shares similarities, we use a common notation and only define terms once unless significant differences exist between different methods.

\paragraph{Simclr} \cite{chen2020simple} proposes a technique to maximize the agreement between representations of two stochastically augmented views of the same image. Initially, the image $i$ is given two views using the stochastic augmentation function $f_{aug}(\cdot)$. Set of augmentations include rotations, flips, color jittering, cropping an image and resizing to its original size. The encoder network $f_{\theta}(\cdot)$ (Resnet50 for this work) with parameters $\theta$ and an auxiliary projection layer $p_{\hat{\theta}}(\cdot)$ with parameters $\hat{\theta}$ is used to obtain projected representations $\bm{z_m}, \bm{z_n} = p_{\hat{\theta}}(f_{\theta}(f_{aug}(i)))$, where $\bm{z_m}$ and $\bm{z_n}$ are different vectors ($\in \mathbb{R}^{128}$) due to the stochastic augmentation function $f_{aug}(\cdot)$. The difference between $\ell_2$ normalized feature representations of the two views per image ($\bm{z_m}$ and $\bm{z_n}$) are minimized while representations of other images in the same training batch are maximized from the image $i$ through a contrastive loss function called \textit{NT-Xent} (the normalized
temperature-scaled cross-entropy loss), defined as \begin{equation}
    \ell_{m,n}= -\log \frac{\exp(\textrm{similarity}(\bm{z_m},\bm{z_n})/\tau)}{\sum_{k=1}^{2N}\mathbbm{1}_{[k\neq i]}\exp(\textrm{similarity}(\bm{z_m},\bm{z_k})/\tau)}, 
\label{eqn:nt_xent}\end{equation} where $\tau$ is the temperature parameter, $N$ is the batch size during pretraining, $\mathbbm{1}$ is the indicator function, and the similarity function is the cosine similarity defined as $\textrm{similarity}(\bm{u},\bm{v}) = \bm{u}^T \bm{v} / \left\lVert\bm{u}\right\rVert \left\lVert \bm{v}\right\rVert$. \cite{chen2020simple} shows NT-Xent performs better when used in downstream tasks than alternative contrastive loss functions such as margin \citep{schroff2015facenet} or logistic \citep{mikolov2013efficient} losses. $p_{\hat{\theta}}(\cdot)$ is a single hidden layer MLP that projects the pre-activation layer output into a lower embedding space. Specifically, given $x$, the MLP converts its input $f_{\theta}(f_{aug}(i))$ by applying the function $p_{\hat{\theta}}(x) = W^{(2)}\sigma(W^{(1)}x)$ where, for Resnet50, $W^{(1)} \in \mathbb{R}^{2048\times2048}$, $W^{(2)} \in \mathbb{R}^{2048\times128}$, and $\sigma(\cdot)$ is the rectifying linear unit activation \citep{nair2010rectified}. Comparing $\bm{z_i}$ and $\bm{z_j}$ was found to be more effective in learning representations than directly comparing the pre-activation layer outputs. Once pretraining is complete, the projection layer is discarded and the encoder is used for the downstream task. 

\paragraph{SwaV} \cite{caron2020unsupervised} proposes a modification to the SimCLR framework by comparing clustering assignments for each projection vector as opposed to directly comparing projection vectors, which removes the requirement of pairwise comparisons. Each auxiliary projection is clustered using an online clustering algorithm \citep{cuturi2013sinkhorn} into a fixed number of clusters while maintaining consistency between cluster assignments for the multiple views of the same image. Unlike SimCLR, this technique uses a multi-crop strategy, where in addition to contrasting two same size images (in this work, $224\times224$ pixels), one enforces consistency between smaller crops of the same image ($96\times96$ pixels). The authors find this approach does not add a substantial memory or computational overhead; however, it improves the downstream performance. While the iterative Sinkhorn-Knopp algorithm used for clustering is computationally demanding, this method can work with smaller batch sizes and achieve similar downstream performance with fewer epochs than the aforementioned method.

\paragraph{BYOL} \cite{grill2020bootstrap} proposes a method to learn representations from raw data without relying on negative samples. Two encoders with identical architectures (a non-trainable target and a trainable online network) are used in a feedback loop for preventing learning a collapsed solution. As in previous techniques, two augmented views of the same image are generated, and one is passed through the online network (resulting in $b_m$), while the other is passed through the target (resulting in $b_n$). Differently from the other techniques, the online network output is transformed a second time using a mapping function $q(\cdot)$, and the distance between $q(b_m)$ and $b_n$ is minimized. The target network parameters are then updated using an exponential moving average of the previous online network parameters until a desired number of iterations have been reached. 

\section{Experiments}\label{sec:experiments}

\paragraph{Datasets} We experiment with natural-scene, satellite, and medical images from different modalities. For pretraining, we use a collage of 40,000 histopathology images extracted from 60 publicly available datasets, denoted as ``Histo" \citep{ciga2020self}, Chest-Xray8 \citep{wang2017chestx} (100,000 training images), breast ultrasound (``US") \citep{al2020dataset} (1000 training images), brain tumor MRI \citep{mridataset} (7,500 training images), Satellite \citep{helber2017eurosat} (27,000 training geo-referenced images covering 13 spectral bands), Stanford Online Products (SOP) \citep{oh2016deep} (59,000 training images), Food 101 \citep{bossard14} (101,000 training images), CelebA-HQ \citep{liu2015faceattributes, karras2017progressive} (30,000 training images), Stanford Cars \citep{krause2013collecting} (16,000 training images), and FGVC Aircraft \citep{maji2013fine} (10,000 training images). Each dataset is public and is available from the Tensorflow dataset catalogue \citep{tensorflow2015-whitepaper} except for the medical datasets, which were curated from their respective references. We deliberately included datasets with few images (e.g., ultrasound) to experiment if it is possible to learn from a small amount of unlabeled data. At the validation stage, we use a COVID19 radiography dataset (classifying images into pneumonia, pulmonary opacification, healthy, and COVID19+)  for Xray \citep{chowdhury2020can}, a binary tumor classification dataset in lymph nodes for histology \citep{Veeling2018-qh}, a facial attributes dataset (assigning images different combinations of age, gender, and ethnicity) for CelebA \citep{zhifei2017cvpr}. These datasets were chosen differently from their pretraining counterparts to verify that the target dataset does not have to come from the same distribution as the source to learn useful representations.  All other networks are validated on the same dataset they were pretrained on, where only the training split was used for pretraining. We report the multi-class classification accuracy on each dataset.

\paragraph{Setup} All experiments are conducted with a Resnet50 network. For pretraining, we use SimCLR \citep{chen2020simple}, SWaV \citep{caron2020unsupervised}, and BYOL \citep{grill2020bootstrap}, and compare if two-stage pretraining ($P2X$) is superior to pretraining from scratch ($P1X$). For $P1X$, we only use the target domain training data, and apply each self supervised technique on a randomly initialized Resnet50. For $P2X$, we use the model weights pretrained on the ImageNet dataset using self-supervision (released by the authors of the original works \citep{chen2020simple, caron2020unsupervised, grill2020bootstrap} with Apache 2.0, Creative Commons, and Apache 2.0 licenses, respectively) for a second stage of pretraining on the target domain training data. Each setting was pretrained for 1000 epochs. The number of iterations per epoch is fixed at 100,000 iterations (i.e., backward passes) to fairly compare datasets with different number of pretraining samples. All methods were pretrained using the same hyperparameters, optimizers, and augmentations described in their respective works, regardless of the domain it was pretrained on. In smaller crop size experiments, we use only 96 pixels for SwaV for simplicity, as opposed to the multi-crop strategy proposed in the original work, which is known to improve the performance. We use up to 4 Tesla V100 GPUs with 32 GBs of memory, depending on the memory requirements for each experiment. Unless otherwise stated, we use a batch size of 1024.

For validation, we compare baselines pretrained on ImageNet using SimCLR, SwaV, and BYOL, as well as networks pretrained with $P1X$ and $P2X$ approaches, where each network is fine-tuned with the validation training set samples. For reference, we report downstream task performances with randomly initialized networks (no pretraining) as well as supervised (Sup) initialization, where a Resnet50 is trained on ImageNet dataset with supervision for 1000-class classification. In other words, ``Sup" refers to weights obtained by supervised training on the Imagenet dataset. In contrast, SimCLR, SwaV, and BYOL refer to the weights obtained using self-supervision that are fine-tuned on validation datasets without a second stage of pretraining. We use the Adam optimizer with the learning rate \texttt{3e-4}, and weight decay of \texttt{1e-4} for each experiment. 

\subsection{Overall comparison}

The results are shown in Table \ref{tab:overall_comparison}. We train each setting (P1X and P2X) for 1000 epochs with a tile length of $224\times224$ pixels. Each method is compared within itself, i.e., the method pretrained on ImageNet dataset is compared with pretraining from scratch (P1X) using the same method, and pretraining from already pretrained weights (P2X).

\begin{table*}
\centering
   \caption{Comparison of classification accuracy for each validation dataset for multiple contrastive techniques with different pretraining procedures. For each group separated by a ruler, we first present the baseline results indicated by the name of the self supervised technique, obtained by applying the pretrained weights (pretrained on ImageNet dataset) to the target downstream task. The two subsequent rows $P1X$ and $P2X$ use the same self supervised method, but use only the target domain data for pretraining. $P1X$ starts pretraining from randomly initialized weights, whereas we use the weights pretrained on ImageNet as initialization for $P2X$.}\label{tab:overall_comparison}
    \begin{tabular}{ccccccccccc}
    Method & Histo & Xray & MRI & US & Satellite & SOP & Food & CelebA & Cars & Aircraft \\
    \midrule
      SimCLR & 82.1 & 91.4 & 79.9 & 82.5 & 91.7 & \textbf{66.8} & 46.6 & 74.2 & 52.0 & 73.9 \\
      P1X & 87.2 & \textbf{94.9} & 76.9 & 67.6 & 91.8 & 49.7 & 36.2 & 76.8 & 44.4 & 50.1 \\
      P2X & \textbf{89.3} & \textbf{94.9} & \textbf{81.2} & \textbf{87.4} & \textbf{95.5} & 64.0 & \textbf{48.6} & \textbf{77.8} & \textbf{67.6} & \textbf{82.3} \\

    \midrule
      SwaV & \textbf{86.9} & 94.2 & 79.2 & 78.1 & 93.2 & 63.8 & 48.5 & 74.5 & 56.8 & 77.0 \\
      P1X & 86.0&93.6&79.4&75.1&93.2&42.4&26.2&73.6&46.8&64.3 \\
      P2X & 85.0 & \textbf{95.6} & \textbf{81.5} & \textbf{88.2} & \textbf{95.4} & \textbf{69.5} & \textbf{56.7} & \textbf{76.2} & \textbf{64.8} & \textbf{80.7} \\

    \midrule
      BYOL & 85.3 & 95.2 & \textbf{83.0} & \textbf{87.4} & 94.6 & 65.6 & 52.8 & 75.8 & 60.8 & \textbf{77.7} \\
      P1X & 76.8 & 90.6 & 75.9 & 64.2 & 79.8 & 34.4 & 14.5 & 69.2 & 10.0 & 39.5 \\
      P2X & \textbf{89.0} & \textbf{95.4} & 82.7 & 85.0 & \textbf{95.4} & \textbf{66.2} & \textbf{53.4} & \textbf{76.3} & \textbf{63.6} & 77.5 \\
    \midrule
      Random & 79.5 & 90.7 & 75.4 & 65.8 & 80.0 & 32.3 & 14.7 & 70.1 & 9.5 & 40.3 \\

      \midrule
      
      Sup & 84.1 & 95.0 & 82.0 & 85.6 & 94.9 & 66.9 & 54.0 & 75.3 & 69.5 & 80.2 \\

    \bottomrule
    \end{tabular}
\end{table*}

\subsection{Low data regime performance on downstream tasks}

An important and attractive property of self-supervised techniques is their ability to perform significantly better under low data regimes compared to randomly initialized networks \citep{henaff2020data, grill2020bootstrap, caron2020unsupervised}. We use 10\% of each validation training set and compare different approaches to understand if further benefits are achievable with the second stage of pretraining on the domain of interest. Each experiment is run thrice with a different 10\% of the same dataset to avoid selecting a portion that favors any specific method. We report the average accuracy over three runs. The results are shown in Table \ref{tab:tenpercent_comparison}. 

\begin{table*}
\centering
   \caption{Comparison of classification accuracy for each validation dataset for multiple contrastive techniques with different pretraining procedures, when only 10\% of the available training data is used. Each experiment is run three times and the average is reported.}\label{tab:tenpercent_comparison}
    \begin{tabular}{ccccccccccc}
    Method & Histo & Xray & MRI & US & Satellite & SOP & Food & CelebA & Cars & Aircraft \\
    \midrule
      SimCLR & 81.2 & 84.0 & 65.7 & 72.5 & 86.6 & 51.2 & 17.3 & 57.9 & 5.2 & 27.4 \\
      P1X & 79.4 & 84.0 & 59.4 & 60.2 & 91.8 & 49.7 & 5.2 & \textbf{65.6} & 4.3 & 8.0 \\
      P2X & \textbf{82.6} & \textbf{90.8} & \textbf{70.1} & \textbf{76.7} & \textbf{92.3} & \textbf{52.6} & \textbf{21.2} & 62.8 & \textbf{5.9} & \textbf{36.1} \\
    \midrule
      SwaV & 82.0 & 88.5 & 66.2 & 70.3 & 84.9 & 41.0 & 11.8 & 60.3 & 3.3 & 23.1 \\
      P1X & 83.1&87.6&70.1&60.4&87.5&27.0&8.0&61.1&4.2&20.0 \\
      P2X & \textbf{85.5} & \textbf{91.6} & \textbf{73.1} & \textbf{74.1} & \textbf{92.0} & \textbf{56.8} & \textbf{28.4} & \textbf{62.7} & \textbf{6.4} & \textbf{33.4} \\

    \midrule
      BYOL & 79.2 & 91.3 & 68.5 & 72.5 & 88.2 & 50.8 & 19.9 & 60.4 & 7.1 & 32.8 \\
      P1X & 73.9 & 88.6 & 48.0 & 57.8 & 62.2 & 17.3 & 3.4 & 46.4 & 2.0 & 7.7 \\
      P2X & \textbf{82.1} & \textbf{91.9} & \textbf{69.1} & \textbf{73.5} & \textbf{88.3} & \textbf{51.9} & \textbf{22.0} & \textbf{61.9} & \textbf{7.3} & \textbf{33.3} \\

    \midrule
      Random & 75.9 & 77.6 & 46.4 & 55.6 & 60.5 & 18.4 & 2.9 & 46.3 & 1.8 & 7.9 \\

      \midrule
      Sup & 82.7 & 92.3 & 71.3 & 71.9 & 89.1 & 55.8 & 25.4 & 60.2 & 7.8 & 33.1 \\

    \bottomrule
    \end{tabular}
\end{table*}

\subsection{The impact of double pretraining on convergence speed}

\begin{figure*}
     \centering
     \includegraphics[width=1\textwidth]{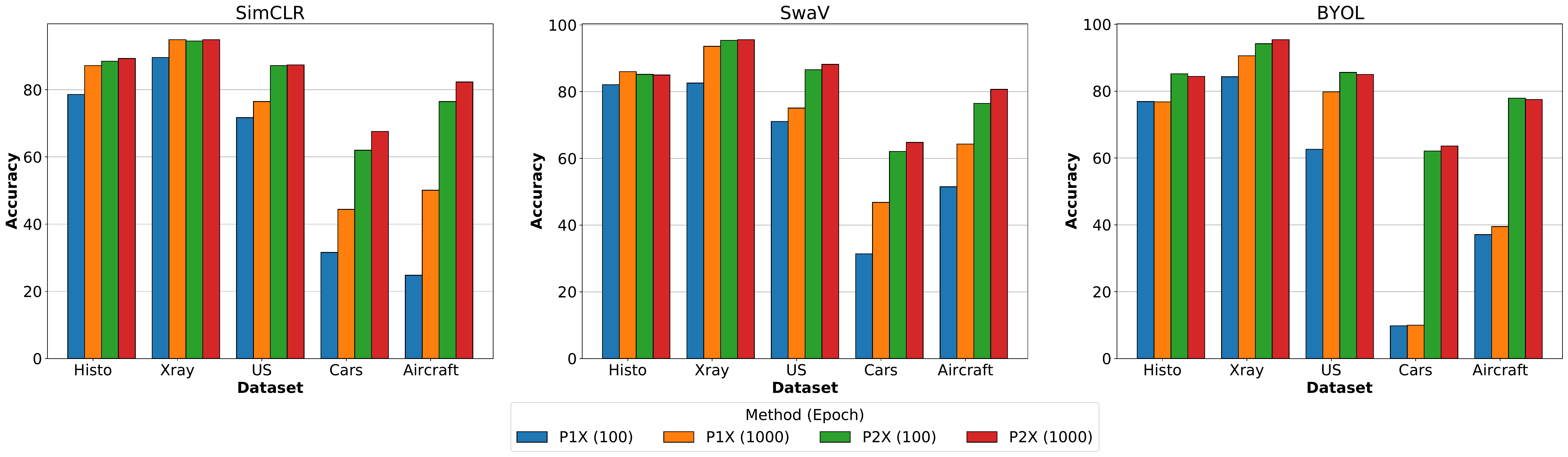}
    \caption{Comparison of downstream performance when the network is pretrained for 100 epochs versus 1000, for P1X (pretraining from scratch) and P2X (pretraining an already pretrained network).}\label{fig:barplot_numepoch}
\end{figure*}

In supervised learning, researchers have previously reported using pretrained initialization leads to faster convergence, but with a trade-off of lower final accuracy \citep{Liu2017Detecting}. We examine if using pretrained networks in the two-step pretraining approach exhibits the same problem. We randomly select three medical and two natural-scene image validation datasets and compare validation accuracy on each method when one pretrains for only 100 epochs vs. 1000 epochs. The results are shown in Fig. \ref{fig:barplot_numepoch}.

\section{Discussion}


We found that the two-step pretraining can improve classification performance for multiple domains, shorten the pretraining duration, and allow for pretraining with smaller batch sizes and smaller images. Furthermore, we found that for some datasets, the two-step approach significantly outperforms pretraining from scratch regardless of the number of epochs the network is pretrained, indicating either the limitations of the current contrastive techniques or that certain datasets are not conducive to learning a rich set of features that can be used for the downstream tasks. Furthermore, we find two-step pretraining is especially useful when only a fraction of the training data is used. One may achieve near-peak accuracy at only 100 epochs when the P2X procedure is applied; therefore, two-step pretraining can be considered as an inexpensive strategy for boosting performance that surpasses both self-supervision and ImageNet pretraining (denoted as ``Sup" in our experiments).

We found that the pretraining converges faster for P2X, which is reflected in the results from Fig. \ref{fig:barplot_numepoch}, where pretraining for 100 epochs versus 1000 epochs has led to a minor improvement compared to P1X for most datasets. We also found that the performance gains reported by the self-supervised methods investigated in this paper do not necessarily hold for smaller datasets from different domains. For instance, while SwaV performs better than SimCLR when pretrained on ImageNet, it was not superior to SimCLR in our single-step pretraining (P1X) experiments.

Overall, we find transferring pretrained features to be an effective strategy to bypass the factors limiting the widespread adoption of contrastive self-supervised learning under the low-resource setting. Our results highlight that while using larger batches for pretraining from scratch may be necessary, fine-tuning these features for the domain of interest does not require large batch training. Moreover, starting from an already pretrained baseline leads to significantly faster convergence with better final downstream task performance.

\section{Conclusion}

In this work, we proposed a simple transfer learning approach based on fine-tuning an already pretrained network to improve downstream task performance. The method we present assumes a network pretrained on a larger dataset with rich visual diversity (e.g., the ImageNet dataset) can be used to improve task performances on a domain of interest that may not exhibit such diversity. Particularly, this method aims to ameliorate the limitations of contrastive objective when the contrasted images lead to learning representations that may not be optimal for the downstream task. As most classification tasks only consist of images from a narrow subset of the visual domain (e.g., medical image analysis on a specific modality or brand identification from car images), using an already pretrained network as initialization which may have encoded more generalized features can perform better on the downstream task compared to pretraining with images only from the domain of interest. 

While most current methods are pretrained on ImageNet dataset, which is limited to natural-scene images, it is possible to pretrain these methods on much larger datasets that cover a wider range of domains (e.g., natural-scene beyond what is represented in ImageNet, medical images, moving images such as movie snapshots). Once these pretrained networks are available, their weights can be used as initialization for further pretraining with the procedure described in this paper. Moreover, it is reasonable to assume this procedure can be applied to future techniques that improve the state-of-the-art.

We believe that the transfer learning approach can also help democratize the efforts in unsupervised learning. While most practitioners do not have access to expensive hardware such as multiple GPUs or TPUs, the method presented here allows for pretraining on GPUs with smaller memory and shorter training times. Furthermore, we have shown that even with very small datasets (e.g., the ultrasound dataset with 780 images), two-step pretraining can improve the downstream task performance. In cases where even unlabeled images are scarce (e.g., medical images that are subject to regulations and ethics board approvals), data efficiency becomes an attractive property of the proposed method.

Finally, the mitigation of labeled data requirements is an important and necessary milestone for the adoption of machine learning into practice, especially when labeling efforts cannot be easily crowdsourced for applications that require expert annotations, such as medical imaging. In such applications, even incremental improvements (e.g., a few percentage points for classification) without additional data can help expedite clinical adoption. We have shown that pretrained networks can be used to achieve various degrees of gains over pretraining from scratch. Furthermore, we have shown that the benefits of two-step pretraining are more pronounced compared to single-step pretraining when only a fraction of the labeled data is used.

\acks{This work was funded by Canadian Cancer Society (grant \#705772) and NSERC. It was also enabled in part by support provided by Compute Canada (www.computecanada.ca).}

\bibliography{jmlr-sample}

\appendix

\section{Additional experiments}

In this section, we conduct two additional experiments to show the efficacy of our approach under low resource settings. The results presented in tables \ref{tab:batch_comparison} and \ref{tab:smallimage_comparison} are relative figures compared to the results presented in Table \ref{tab:overall_comparison}, Section \ref{sec:experiments}.

\subsection{The impact of batch size}

Most contrastive techniques require large batch training or memory banks to learn representations. This section examines if using pretrained weights as initialization for self-supervision can mitigate these requirements. As opposed to our original experiments, where the batch size for training was set to be 1024, we use a batch size of 32 and pretrain P1X and P2X approaches as described before. We chose a batch size of 32 as it was the largest value ($2^x$ where $x$ is an integer) that could fit a GPU with 8 GBs of memory with the most demanding method investigated in this work (SwaV). The results shown in Table \ref{tab:batch_comparison} indicate the difference in accuracy compared to the corresponding entries in Table \ref{tab:overall_comparison}. For instance, $-4.0$ for the Simclr $\Delta$ P1X Histo entry indicates the accuracy drop when the models are pretrained using a batch size of 32. Since the value in Table \ref{tab:batch_comparison} is 79.5, the accuracy for this entry is 75.4. Conversely, if the accuracy has improved, the difference is positive. When comparing P1X vs. P2X, the more positive value is considered better, regardless of the initial accuracy when one pretrains with a batch size of 1024.

We found that using a smaller batch size has a more pronounced negative impact on the P1X compared to P2X for most settings (Table \ref{tab:batch_comparison}). In few settings where P1X achieves a slightly better result compared to P2X (e.g., for SimCLR, MRI and Aircraft datasets, or for SwaV, CelebA dataset), the differences are comparable (within 3 to 4\%). In contrast, the degradation gap between two schemes  (P1X and P2X) is relatively high for others (e.g., Cars dataset for SimCLR and SwaV, or SOP and Aircraft dataset for SwaV). Therefore, we conclude P2X is more stable for settings where the pretraining is done with a small number of images. Finally, the differences between the two schemes are not as significant for BYOL for most validation datasets. BYOL does not require negative pairs for self-supervision; therefore, BYOL is impacted less when fewer pretraining samples per iteration are used.

\begin{table*}
\centering
   \caption{Difference in accuracy on downstream tasks for three methods using P1X and P2X procedures when the pretraining was done with a batch size of 32. The values represent the drop or gain in accuracy with respect to the corresponding entries in Table \ref{tab:overall_comparison}, where the batch size was 1024.}\label{tab:batch_comparison}
    \begin{tabular}{ccccccccccc}
    Method & Histo & Xray & MRI & US & Satellite & SOP & Food & CelebA & Cars & Aircraft \\
    \midrule
      \textbf{Simclr} &  &  &  & & &  & &  &  & \\

     $\Delta$ P1X & -4.0 & -3.3 & \textbf{+2.5} & -2.9 & -8.6 & -0.2 & -18.0 & -4.9 & -36.4 & \textbf{-8.4} \\

      $\Delta$ P2X & \textbf{-3.1} & \textbf{0.0 }& -1.8 & \textbf{-1.6} & \textbf{-1.2} & \textbf{+2.5} & \textbf{-3.2} & \textbf{-2.0} & \textbf{-18.7} & -10.8 \\

    \midrule
      \textbf{SwaV} &  &  &  & & &  & &  &  & \\

      $\Delta$ P1X & -1.3&-1.7&-0.2&-6.4&-8.6&-6.3&-5.7&\textbf{-0.6}&-23.1&-13.0\\
      $\Delta$ P2X & \textbf{+3.4}&\textbf{+0.1}&\textbf{+0.5}&\textbf{-4.2}&\textbf{+0.3}&\textbf{+1.3}&\textbf{-2.0}&-1.9&\textbf{-6.6}&\textbf{+1.1} \\

    \midrule
      \textbf{BYOL} &  &  &  & & &  & &  &  & \\

      $\Delta$ P1X & +0.5 & \textbf{+0.6} & -1.0 & -0.3 & \textbf{+2.2} & \textbf{-0.2} & -0.7 & \textbf{+1.2} & -1.5 & \textbf{-0.6} \\

      $\Delta$ P2X & \textbf{+2.6} & 0.0 & \textbf{0.0} & \textbf{+1.1} & +0.9 & -0.4 & \textbf{+0.6} & -2.2 & \textbf{+1.5} & -1.2 \\

    \bottomrule
    \end{tabular}
\end{table*}

\subsection{Pretraining with smaller images}

Time and memory requirements for pretraining with images of size $224 \times 224$ pixels are significantly more than for images of size $96 \times 96$ pixels. For the former, one may fit 128 images (per batch) on a 32 GB memory when pretraining a Resnet50 with the SimCLR method, whereas 350 images can be fit for the latter. We found that pretraining with larger images take three times longer for the three methods we investigated in this paper, when Tesla V100 GPUs are used. Using smaller images allows for larger batch sizes as well as shorter training times; however is only beneficial if the difference in downstream task accuracy is negligible.

Interestingly, we found that the downstream accuracy did not deteriorate when we pretrained with downsampled images (i.e., $96\times96$ pixels instead of $224\times224$) for both P1X and P2X. Although we did not observe a superiority of P2X over P1X, we report results on Table \ref{tab:smallimage_comparison}, as we believe these findings may be of interest to researchers due to the possible memory and time savings, when either method is used.

\begin{table*}
\centering
   \caption{Difference in accuracy on downstream tasks for three methods using P1X and P2X procedures when the pretraining was done with an image size of $96 \times 96$. The values represent the drop or gain in accuracy with respect to the corresponding entries in Table \ref{tab:overall_comparison}, where the image size was $224 \times 224$.}\label{tab:smallimage_comparison}
    \begin{tabular}{ccccccccccc}
    Method & Histo & Xray & MRI & US & Satellite & SOP & Food & CelebA & Cars & Aircraft \\
    \midrule
      \textbf{Simclr} &  &  &  & & &  & &  &  & \\

     
     $\Delta$ P1X & \textbf{-3.8} & -2.4 & \textbf{+2.5} & \textbf{+6.7} & \textbf{-1.0} & -3.1 & -4.7 & -1.6 & -32.0 & \textbf{+0.6} \\


     $\Delta$ P2X & -4.9 & \textbf{+0.1} & +1.0 & -3.2 & \textbf{-1.0} & \textbf{+3.3} & \textbf{+4.4} & \textbf{-1.4} & \textbf{-5.0} & -3.5 \\

    \midrule
      \textbf{SwaV} &  &  &  & & &  & &  &  & \\

      $\Delta$ P1X & -1.9&-0.9&\textbf{+2.1}&-6.9&-11.8&\textbf{+3.2}&-10.6&-4.3&-36.3&-12.8 \\
      $\Delta$ P2X & \textbf{+1.4}&\textbf{-0.5}&+1.0&\textbf{-5.0}&\textbf{+0.1}&-0.2&\textbf{-1.3}&\textbf{-0.8}&\textbf{+0.3}&\textbf{-2.4}\\

    \midrule
      \textbf{BYOL} &  &  &  & & &  & & &  & \\

      $\Delta$ P1X & \textbf{+1.7} & \textbf{+1.1} & \textbf{+1.3} & \textbf{+2.1} & \textbf{+0.8} & -2.6 & -0.5 & \textbf{+1.0} & \textbf{0.0} & \textbf{+15.3} \\
      
     $\Delta$ P2X & -4.6 & -0.4 & -0.5 & -0.8 & -0.9 & \textbf{+1.1} & \textbf{-0.4} & +0.1 & -1.0 & +1.3 \\

    \bottomrule
    \end{tabular}
\end{table*}



\end{document}